\documentclass[aps,superscriptaddress,prb,twocolumn]{revtex4}
\usepackage{graphicx}
\usepackage{amsmath}
\usepackage{amssymb}
\usepackage{color}
\usepackage[hidelinks]{hyperref}

\begin{document}

\title{Mode Delocalization in Disordered Photonic Chern Insulator}

\author{Udvas Chattopadhyay}

\affiliation{Division of Physics and Applied Physics, School of Physical and Mathematical Sciences,\\
Nanyang Technological University, Singapore 637371, Singapore}

\author{Sunil Mittal}

\affiliation{Department of Electrical and Computer Engineering, The University of Maryland at College Park, College Park, MD 20742, USA}

\author{Mohammad Hafezi}

\affiliation{Department of Electrical and Computer Engineering, The University of Maryland at College Park, College Park, MD 20742, USA}

\author{Y.~D.~Chong}

\email{yidong@ntu.edu.sg}

\affiliation{Division of Physics and Applied Physics, School of Physical and Mathematical Sciences,\\
Nanyang Technological University, Singapore 637371, Singapore}

\affiliation{Centre for Disruptive Photonic Technologies, Nanyang Technological University, Singapore 637371, Singapore}

\begin{abstract}
In disordered two dimensional Chern insulators, a single bulk extended mode is predicted to exist per band, up to a critical disorder strength; all the other bulk modes are localized.  This behavior contrasts strongly with topologically trivial two-dimensional phases, whose modes all become localized in the presence of disorder.  Using a tight-binding model of a realistic photonic Chern insulator, we show that delocalized bulk eigenstates can be observed in an experimentally realistic setting.  This requires the selective use of resonator losses to suppress topological edge states, and acquiring sufficiently large ensemble sizes using variable resonator detunings.
\end{abstract}

\maketitle
\section{Introduction}

One of the most intriguing features of the Integer Quantum Hall Effect \cite{Klitzing1980, prange_girvin_1990, Wysokinski_2000} is the extraordinary accuracy of quantization in the Hall resistivity of about 1 part in $10^9$.  Disorder plays an important role in this phenomenon; without disorder, the Integer Quantum Hall Effect's celebrated conductance quantization plateaus could not exist \cite{prange_girvin_1990}.  As the seminal work of Anderson and co-workers has shown, the effects of disorder are strongly dependent on the spatial dimensionality \cite{Anderson1958, Abrahams1979, Evers2008Rev}. In one dimension, arbitrarily weak disorder localizes all states, whereas three dimensional systems host localized states at low energies and extended states at high energies, separated by a mobility edge.  In two dimensions (2D), the effects of disorder depend on the time-reversal and spin symmetries of the system \cite{Garcia2013}. For normal 2D materials, which are in the orthogonal symmetry class, all states are localized by disorder, similar to the one dimensional case \cite{MacKinnon1981, MacKinnon1983}.  For the unitary class, which includes Integer Quantum Hall systems and other Chern insulators, localization occurs via a mechanism called ``levitation and annihilation'': in the limit of infinite system size, the introduction of disorder causes all states to localize except for one state per (topologically nontrivial) band, which remains extended; with increasing disorder strength, extended states in adjacent bands can move towards each other and annihilate, producing a transition to a purely-localized phase \cite{Khmelnitskii1984, Laughlin1984, Evangelou1995, Tomi2002, Onoda_QSH2007}.

The inter-plateau longitudinal conductance peaks in the Integer Quantum Hall Effect constitute the principal experimental evidence for the special bulk extended states in 2D unitary disordered systems \cite{Mares1999, Coleridge2005}.  In Chern insulators without Landau levels, there is thus far little experimental evidence for these states, nor for theorized behaviors such as levitation and annihilation, though many numerical studies have been performed \cite{Evangelou1995, Tomi2002, Onoda_QSH2007, Xu2012, Castro2015, Qiao2016}.  In condensed matter settings, such experiments are very challenging due to the need to fabricate large samples with controlled amounts of disorder.

This paper investigates the possibility of using photonics to probe the localization behavior of 2D Chern insulators.  Over the last decade, photonics has emerged as a versatile setting for realizing phenomena associated with band topology \cite{Ozawa2019, Kim2020}, including reflectionless edge transport \cite{Raghu2008, Wang2009, rechtsman2013, Hafezi2011, Hafezi2013, Daniel2018, Mittal2019}, topological pumping \cite{Zilberberg2012, Hu2015, Hu2017, Mittal2016, Zilberberg2018}, spin and valley Hall edge states\cite{Wu-Hu, Ma2016}, Fermi arcs \cite{Lu2013}, and more.  Topological phenomena also hold promise for novel device applications in photonics, such as highly-robust waveguides and delay lines \cite{Wang2009, Hafezi2011, Hafezi2013}, amplifiers \cite{Peano2016}, isolators \cite{Zhou2017}, and lasers \cite{Harari2018, Bandres2018, Zeng2020}.  There are several reasons to consider using photonic topological insulators to study the localization properties of topological phases.  First, different disorder configurations can be implemented on a single device by means of optical, thermal, acoustic, or electrical pumps \cite{Villeneuve1996, Mittal2016, Daniel2018}, which should simplify the acquisition of ensembles with many independent disorder realizations. Second, it is possible to excite any frequency in the band or band gap via a number of available launching schemes. Third, field distributions can be observed by near-field imaging or other techniques, allowing for the accurate and direct determination of quantities such as localization lengths \cite{Hafezi2013, Mittal2019}.  Fourth, losses can be controllably incorporated into photonic structures \cite{ElGanainy2019}, which, as we shall see, is helpful for distinguishing the experimental signatures of bulk delocalization from the effects of topological edge states.  Although photonics has already been extensively employed for the experimental study of ordinary Anderson localization \cite{segev2013}, it has never been used to investigate the peculiar localization properties of bulk states in Chern insulators.

The downside, however, is that photonic Chern insulators must be formed from deliberately structured photonic media, such as photonic crystals, metamaterials, coupled resonators, or waveguide arrays \cite{Ozawa2019, Kim2020}.  In some of these platforms, fabrication technologies are unable to create lattices that are sufficiently large, relative to the unit cell, for localization studies.  Moreover, several platforms exhibit rather high radiative and material losses; while a loss-induced decay length of (say) several unit cells might be acceptable for the purposes of demonstrating topological edge transport, it could complicate localization studies through the introduction of exponentially decreasing intensity profiles to states that are supposed to be delocalized.

We focus on a type of 2D photonic Chern insulator consisting of a lattice of on-chip coupled ring resonators, which has recently been proposed and implemented \cite{Daniel2018, Mittal2019}.  This system is amenable to theoretical analysis since it can be accurately described by a tight-binding model \cite{Daniel2018}.  It has been experimentally realized using silicon photonics, featuring large lattice sizes of up to $15\times7$ unit cells and sufficiently low levels of loss that edge transport was observed over distances of over a dozen unit cells \cite{Mittal2019}.  The lattice parameters can be altered by methods such as optical pumping \cite{Daniel2018, Mittal2019}, so disorder ensembles can be readily generated in each photonic lattice through spatially inhomogenous pumping, eliminating the need to fabricate many different samples.
\begin{figure}
  \centering
  \includegraphics[width=0.475\textwidth]{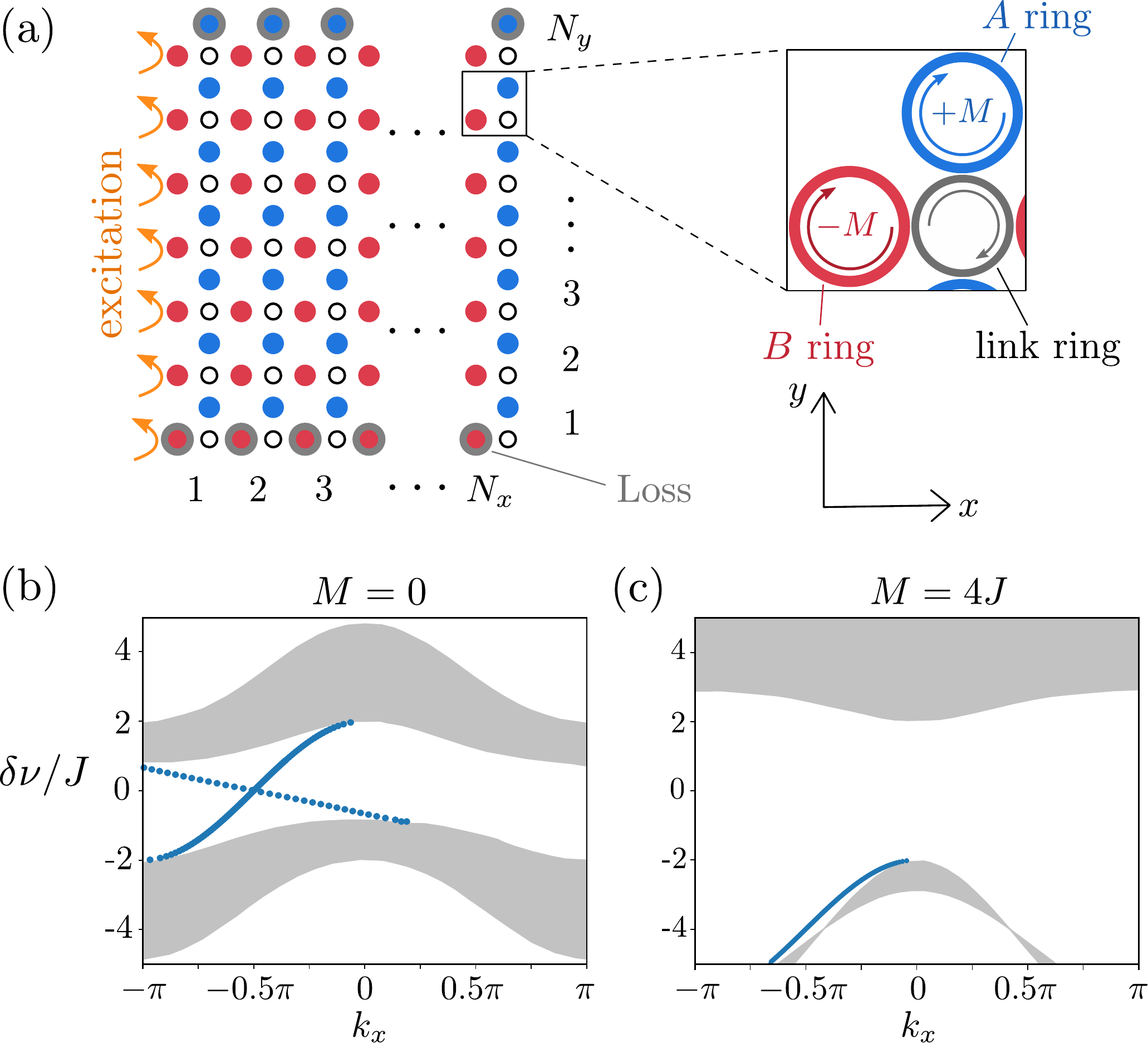}
  \caption{(a) Schematic of a finite lattice of coupled optical resonators, composed of $N_x$ unit cells along $x$ (length) and $N_y$ unit cells along $y$ (width).  A loss term $i \gamma$ is added to the sites along the top and bottom edges. Light is injected into the sites on the first column via uniformly excited coupling waveguides (orange arrows); there are no coupling waveguides on the rightmost column, which serves as a closed edge.  Inset: close-up view of one unit cell, showing the selected direction of circulation within the resonators.   (b)--(c) Calculated band structures of the semi-infinite lattice (infinite length and finite width, with losses omitted), for (b) the topological phase ($M=0$) and (c) the trivial phase ($M=4J$). Edge states are shown in blue. In (b), the right-moving (left-moving) edge state is localized on the bottom (top) edge. In (c), there are edge states localized on the bottom edge, but these do not span the gap.  }
  \label{Device_BandStructure}
\end{figure} 

Using tight-binding simulations, we show that this platform could be used to access the delocalization of bulk Chern insulator states and the levitation and annihilation phenomenon.  A clear experimental signature can be achieved with a lattice size of about $50 \times 12$, which is a modest increase relative to existing experiments \cite{Mittal2019}, and far smaller than the lattices in previous numerical localization studies (which typically feature sample lengths of $~10^3$ or more) \cite{MacKinnon1981, MacKinnon1983, Evangelou1995, Tomi2002, Wang2015}.  Silicon-on-insulator, which is typically used in experiments involving coupled ring resonators \cite{Mittal2019}, has a loss level of $\sim 1 \, \textrm{dB/cm}$. While silicon is the best choice for compactness, it is not particularly a low-loss platform. Silicon nitride can achieve loss levels of $\sim 1 \, \textrm{dB/m}$ \cite{Bauters2011} and we show that such a level of loss does not affect the key results.  The topological edge states of the Chern insulator tend to conflict with the experimental signature of the delocalized bulk states, but we find that the former can be suppressed simply by adding losses to the resonators along the lattice edges.  Hence, the photonic lattice can provide a way to explore the localization behavior of bulk states in disordered topological insulators, which have thus far resisted in-depth experimental investigation.

\section{Model}

We consider a photonic Chern insulator of a type that has recently been proposed \cite{Daniel2018} and implemented using silicon photonics \cite{Mittal2019}.  As shown in Fig.~\ref{Device_BandStructure}(a), the system consists of an bipartite square lattice of resonant ``site rings'' coupled to off-resonant ``link rings''.  The site rings occupy two sublattices, denoted by $A$ and $B$, and the link rings introduce nearest-neighbor and next-nearest-neighbor couplings between them.  Light propagation within the lattice can be decomposed into two pseudospin sectors (corresponding to clockwise or counterclockwise circulation in the site rings), which do not interact due to local momentum conservation at the inter-ring coupling regions \cite{Hafezi2011, Hafezi2013}.  Within each sector, time reversal symmetry is effectively broken (however, the physical structure is time reversal symmetric and can be fabricated from ordinary dielectric materials).

In the absence of disorder and losses, the system is described by the following tight-binding Hamiltonian \cite{Daniel2018}:
\begin{equation}
  H = (\epsilon + M) \, n_A + (\epsilon - M)\, n_B + V_{\textrm{nn}} + V_{\textrm{nnn}},
\end{equation}
where
\begin{align}
  \epsilon &= 2J \cot(\phi/2) \\
  n_{A} &= \sum_{r} a_{r}^\dagger \,a_{r}, \;\;\;
  n_{B} = \sum_{r} b_{r}^\dagger \,b_{r} \\
  V_{\textrm{nn}} &=  W_1 \sum_r \left[a_{r}^\dagger(b_{r} + b_{r+x+y}) + b_{r}^\dagger(a_{r-x}+a_{r-y}) \right] \\
  V_{\textrm{nnn}} &= W_2 \sum_r \Big[ a_{r}^\dagger
    \sum_{\pm}\hat{a}_{r\pm y} + b_{r}^\dagger \sum_{\pm} b_{r\pm x} \Big] \\
  W_1 &= J \exp (i\phi/4) \csc (\phi / 2), \;\;\;
  W_2 = J \csc(\phi/2).
\end{align}
Here, $a_{r}$ and $b_{r}$ are the annihilation operators for the $A$ and $B$ sublattices on the unit cell at position $r$ (with $r+x$ denoting the position one unit cell to the right, etc.), $M$ is a sublattice-dependent resonator detuning for the site rings, and $J$ and $\phi$ parameterize the couplings mediated by the link rings.  For $\phi \ne 0$, time reversal symmetry is effectively broken.  Each eigenvalue of $H$, denoted by $\delta \nu$, corresponds to the detuning of a photonic eigenmode relative to a reference frequency.

\begin{figure*}
  \centering
  \includegraphics[width=0.99\textwidth]{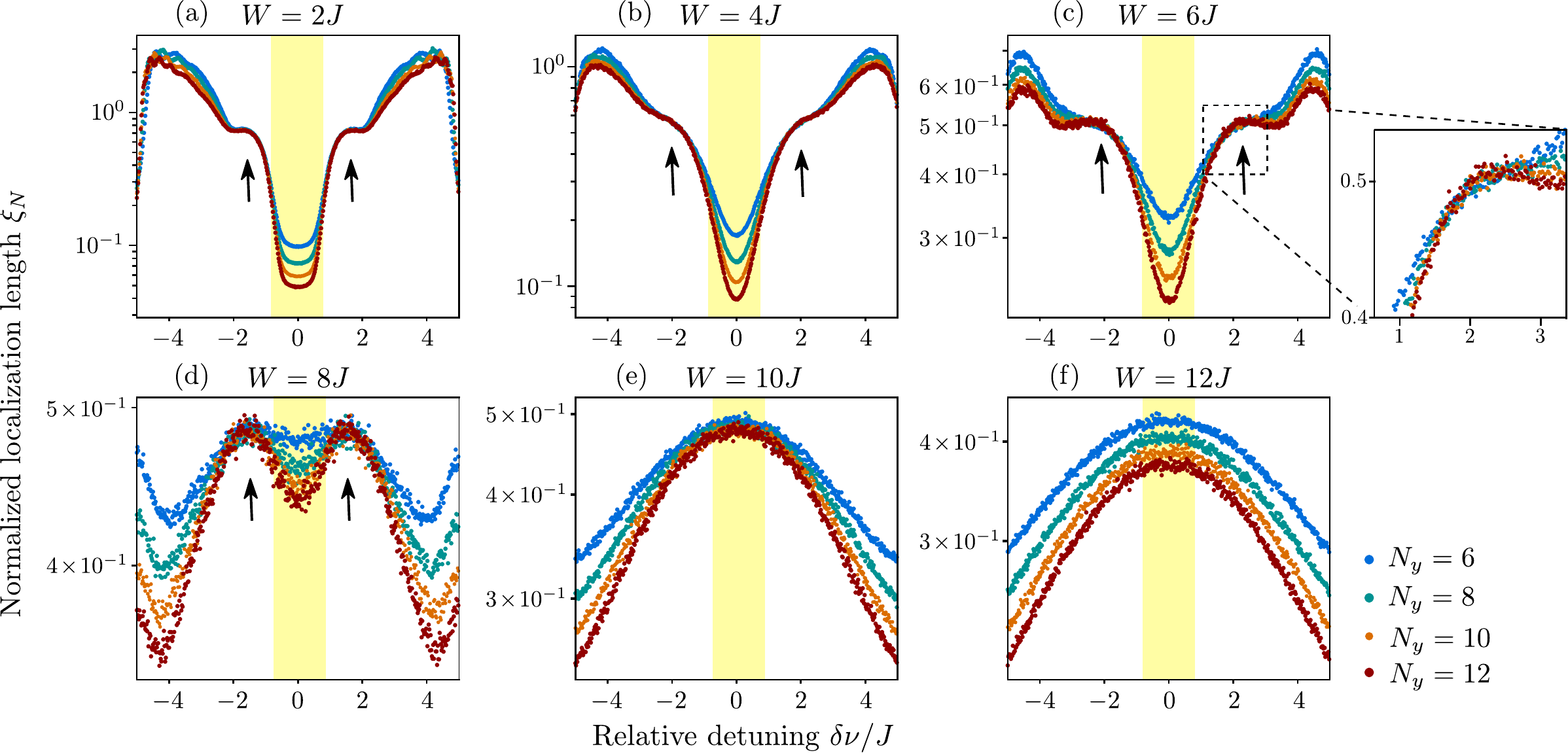}
  \caption{Normalized localization length $\xi_N$ versus relative detuning $\delta \nu$ for samples with length $N_x = 50$, different widths $N_y$, different disorder strengths $W$, and open boundary condition
along $y$. A loss term $-i \gamma$ with $\gamma=J$ is added to the edge sites to suppress the edge state.  The ensemble size used is $500$. Black arrows indicate the critical behavior ($\xi_N$ increasing or constant with $N_y$), as shown in the inset for $W=6$. For large $W$, levitation and annihilation of the critical states is observed, and the bulk states are completely localized for $W \geq 10J$. The band gap of the ordered bulk system is highlighted in yellow.}
  \label{Main_HBC}
\end{figure*} 

For a given nonzero $\phi$, which depends on the detuning of the link rings relative to the site rings, the lattice supports both a topological band insulator phase (a Chern insulator) and a topologically trivial phase (a normal insulator), depending on the value of $M/J$, the relative detuning between the site ring sublattices \cite{Daniel2018}.  In the following, we take $\phi = \pi$ (i.e., link rings exactly anti-resonant with the site rings), with $M = 0$ for the Chern insulator and $M = 4J$ for the normal insulator.  In Fig.~\ref{Device_BandStructure}(b)--(c), their bandstructures are plotted for a disorder-free quasi-one-dimensional geometry (i.e., a strip that is infinite in $x$, and finite in $y$ with open boundary conditions).  For the Chern insulator ($M=0$), the gap is spanned by chiral edge states that are localized to opposite edges, consistent with the Chern numbers of $\pm 1$ possessed by the two bulk bands \cite{Ozawa2019, Kim2020, Daniel2018, Mittal2019}.  For the normal insulator ($M = 4J$), there are no edge states spanning the gap.

We consider rectangular lattices of length $N_x$ and width $N_y$.  As indicated in Fig.~\ref{Device_BandStructure}(a), light is injected uniformly into the lattice via waveguides coupled to the site rings on the left edge (i.e., all site rings along that edge are excited with equal intensity and phase).  The light returning from the lattice bulk is assumed to outcouple through the same coupling waveguides.

In an actual experiment, the intensities on individual resonators can be determined from direct measurements of weak light scattering \cite{Hafezi2013, Mittal2019}.  The site intensities in column $n$ can be calculated from the frequency-domain Green's function (which we will obtain using the method of Kramer and McKinnon \cite{MacKinnon1983}, which is based on the recursive Green's function technique \cite{Thouless1981, Lewenkopf2013, Zhang2017}).  However, the calculation has a subtle dependence on whether the excitations on column $1$ are mutually coherent or incoherent.  First, consider the incoherent case, in which different sites bear no fixed phase relationship with one another.  The total intensity on column $n$ is $\mathrm{Tr}\left(\mathcal{G}^\dagger\mathcal{G}\right)$, where $\mathcal{G}$ is the Green's function matrix beween sites on column $1$ and sites on column $n$.  We can thus define
\begin{equation}
  T = \frac{1}{N_y}\, \mathrm{Tr}\left(\mathcal{G}^\dagger\mathcal{G}\right),
  \label{Tstandard}
\end{equation}
which is the standard definition of the transmittance as used in the electronic transport literature for determining the conductance of a sample \cite{MacKinnon1983}.  On the other hand, in photonic experiments the inputs typically originate from a single laser source with strong spatial coherence.  In that case, the effective transmittance in column $n$ is
\begin{equation}
  T_{c} = \frac{\langle \psi_1 | \mathcal{G}^\dagger\mathcal{G} | \psi_1 \rangle}{\langle \psi_1 | \psi_1 \rangle},
  \label{Tcoherent}
\end{equation}
where $|\psi_1\rangle$ is the input vector (e.g., $[1, 1, \dots, 1]$ if all input waveguides have the same phase).

Anderson-type disorder is introduced into the lattice in the form of a random detuning on each site ring, drawn independently from a uniform distribution over $[-W/2, W/2]$, where $W$ is a tunable disorder strength parameter.  In an actual photonic lattice, such disorder is introduced in part by unavoidable fabrication imperfections; a previous experiment found this to be on the order of the coupling strength $J$ (the basic ``energy'' scale of the tight-binding model) \cite{Mittal2019}.  Additional disorder can be introduced through spatially inhomogenous pumping \cite{Villeneuve1996, Mittal2016, Daniel2018}.  We assume that an ensemble of independent disorder configurations can thus be achieved.

We can then compute the disorder average $\langle \log(T) \rangle$ for the incoherent input case.  This is related to the localization length $\xi$ by \cite{Slevin2001, Paulin2012}
\begin{equation}
  \langle \log (T) \rangle \propto -\frac{2n}{\xi}.
  \label{Trans}
\end{equation}
Therefore, $\xi$ can be extracted from a linear fit of $\langle \log (T)
\rangle$ against $n$.  For coherent inputs, we substitute $T$
with $T_c$ in Eq.~\eqref{Trans}.  We find numerically that although the values of $\langle\log T\rangle$ and $\langle \log T_c\rangle$ are generally different, both cases yield the same fitted value of $\xi$, as shown in Appendix \ref{AppendixA}.  Hence, the use of spatially coherent inputs in a photonic experiment is consistent with the standard definition of the localization length, based on Eq.~\eqref{Tstandard}, which is what our subsequent numerical results are based on.

We then define the normalized localization length \cite{MacKinnon1981}
\begin{equation}
  \xi_N(\delta \nu ,W) = \frac{\xi}{N_y},
  \label{Norm_loc_len}
\end{equation}
which depends implicitly on the operating frequency detuning $\delta\nu$, as well as the disorder strength $W$.  In the localized regime, $\xi_N$ decreases with $N_y$ and vanishes in the $N_y \to \infty$ limit. When extended states are present, $\xi_N$ increases with $N_y$ and diverges in the $N_y \to \infty$ limit. For the critical states at a mobility edge, $\xi_N$ approaches a finite constant in as $N_y \to \infty$. Hence, for given $\delta\nu$ and $W$, we can detect the presence of extended states by finding how $\xi_N(\delta \nu)$ varies with $N_y$ \cite{Xu2012, Evangelou1995, Su_Srep2016, Wang2015, Castro2016}.

In numerical studies of bulk localization in 2D lattices, it has been conventional to take periodic boundary conditions along the upper and lower edges, so that the sample forms an edgeless waveguide \cite{MacKinnon1981, MacKinnon1983, Tomi2002, Wang2015}.  The reason for doing this is that Chern insulators and other 2D topological insulators host topological edge states that are robust against disorder and extended along the edge, which can interfere with the signature of bulk localization of delocalization.  However, since we aim to investigate the feasibility of a realistic on-chip photonic experiment, it is \textit{not} appropriate for us to impose periodic boundary conditions.  Instead, we introduce losses to the sites residing at the edge of the system, as shown by the gray outlines in Fig.~\ref{Device_BandStructure}(a).  The losses are modelled as an imaginary contribution to the detuning, $-i\gamma$, where we take $\gamma = J$ (later, we will also study the effects of smaller losses on all the other sites).  In experiments, losses can be deliberately introduced by using adding lossy materials or claddings or additional scattering defects to the resonators \cite{ElGanainy2019, Gao2016}.  It would be desirable to avoid significantly altering the real detuning of the edge resonators while doing so; otherwise, the edge states may simply be shifted onto adjacent rows further into the bulk.

\section{Results}

Fig.~\ref{Main_HBC} shows the normalized localization length $\xi_N$ versus the source frequency detuning $\delta \nu$, for different disorder strengths $W$ and different lattice widths $N_y$.  The lattices are in the Chern insulator phase ($M = 0$), and have fixed length $N_x=50$.  At the upper and lower edges, we impose open (Dirichlet-like) boundary conditions.  As mentioned, losses are added to the sites on the upper and lower edges to suppress the topological edge state; the losses on the other resonators are assumed to be negligible.

For most values of $\delta \nu$, we find that $\xi_N$ decreases with increasing $N_y$.  As explained in the previous section, this indicates that the eigenmodes at these frequencies are localized by the disorder.  However, in Fig.~\ref{Main_HBC}(a)--(d), representing disorder strengths up to $W = 8J$, there are two regions on each side of the band gap where $\xi_N$ is constant or increases with $N_y$.  This is evidence for the bulk extended states believed to exist in disordered Chern insulators \cite{Evangelou1995, Tomi2002, Onoda_QSH2007, Xu2012, Castro2015, Qiao2016}.  In agreement with theoretical predictions, there is only a narrow range of frequencies within each band where this phenomenon occurs.

The signature of delocalization is more easily observable if the ensemble size is large.  In Fig.~\ref{Main_HBC}, each data point is averaged from an ensemble of $500$ independent disorder realizations.  It would be experimentally unfeasible to fabricate one physical sample for each disorder realization, but it should be possible to implement disorder dynamically via spatially inhomogenous and actively-switchable optical, thermal, acoustic, or electrical pumps \cite{Villeneuve1996, Mittal2016, Daniel2018}.  This would allow numerous independent disorder realizations to be generated with a few physical samples (corresponding to different values of $N_y$).  Furthermore, according to our simulations, delocalization may still be observable for ensemble sizes of as low as $50$.

\begin{figure}
  \centering
  \includegraphics[width=0.475\textwidth]{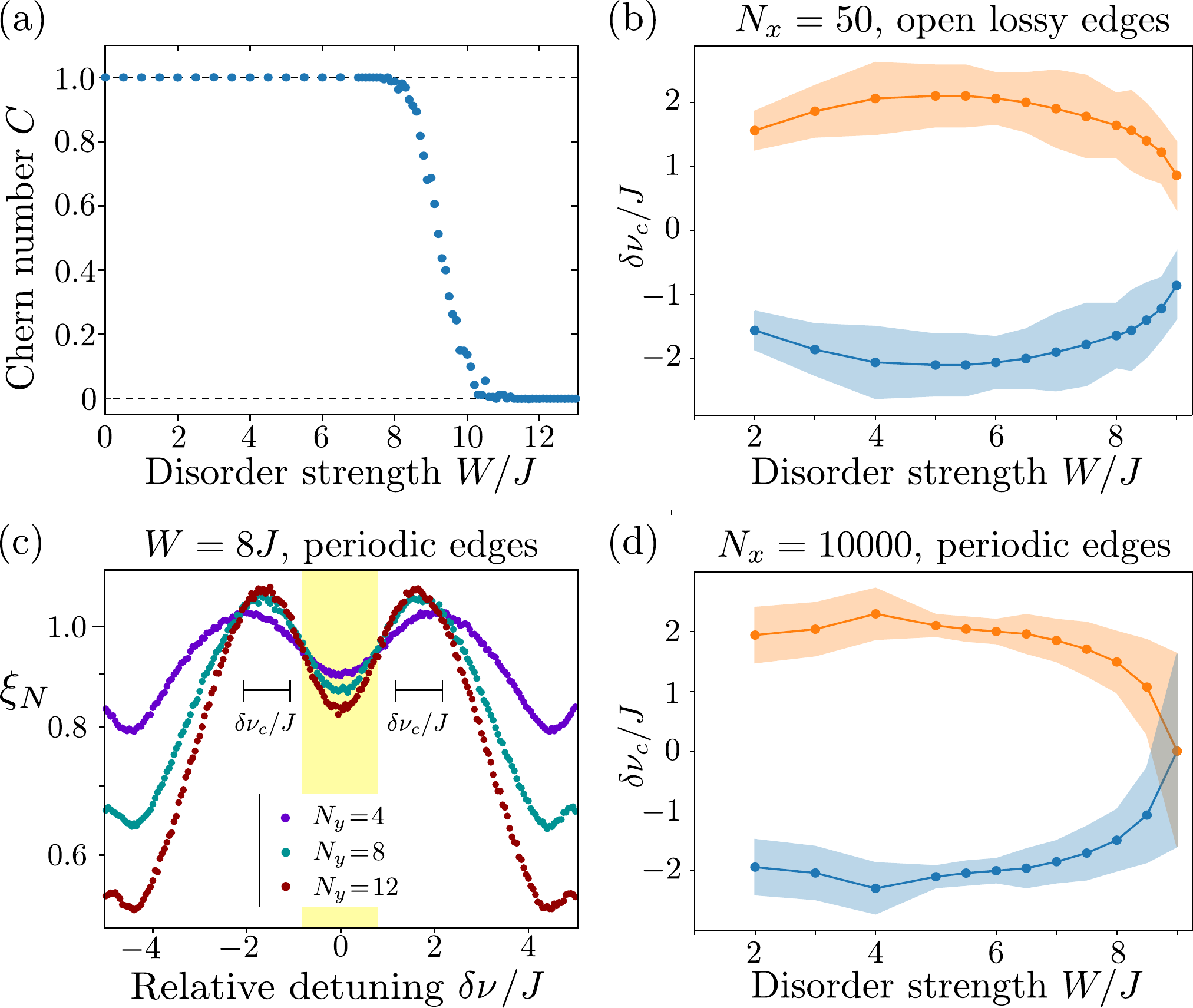}
  \caption{(a) Chern number of the lower band versus disorder strength $W$ for a $32 \times 32$ lattice (averaged over $200$ disorder realizations). (b) Critical energy $\delta \nu_c$ (where the extended state is located) versus $W$ for the realistic sample corresponding to Fig.~\ref{Main_HBC}.  Each shaded region indicates the detuning range over which $\xi_N$ does not decrease with $N_y$, and the dots and solid lines indicate the mid-point of the range.  (c) Normalized localization length versus $\delta \nu$ for the hermitian quasi-1D system ($N_x = 10000$) with periodic boundary along $y$ at a disorder strength $W=8J$. (d) Critical energy $\delta \nu_c$ versus $W$ for semi infinite lattice corresponding to (c). Anderson localization transition by annihilation can be observed at $W=9J$.}
  \label{Fig:3}
\end{figure} 

Fig.~\ref{Fig:3}(a) shows the Chern number calculated for the lower band versus the disorder strength $W$ \cite{Zhang2013, Castro2015}.  The Chern number is computed by summing the Berry flux through each plaquette of the discretized Brillouin zone following Fukui's method \cite{Fukui2005}, using a computationally efficient coupling-matrix method to calculate the Berry flux \cite{Zhang2013, Castro2015}.  With increasing $W$, the Chern number is quantized to unity until $W \approx 8J$, after which it decreases to zero.  This is consistent with the results shown in Fig.~\ref{Main_HBC}, where the signature of the delocalized bulk mode, a property of the Chern insulator, disappears between $W = 8J$ and $W = 10J$.  In Fig.~\ref{Fig:3}(b), we plot the range of critical frequencies (i.e., the frequency range over which $\xi_N$ does not decrease with $N_y$) versus $W$.  As the system transitions from Chern insulator to normal insulator, the critical frequencies shift toward one another, and for $W \gtrsim 9J$ become impossible to distinguish over the statistical noise.  Hence, a photonic lattice can provide evidence for the long-standing theoretical prediction that topologically insulating behavior is destroyed by strong disorder through the levitation and annihilation of the delocalized bulk states \cite{Khmelnitskii1984, Laughlin1984, Evangelou1995, Tomi2002, Onoda_QSH2007}.

In Fig.~\ref{Fig:3}(c), we present simulation results for an idealized and experimentally unrealistic lattice with periodic boundary conditions on the upper and lower edges, and a much greater length of $N_x = 10^4$.  Delocalization is observed at around the same frequencies as in Fig.~\ref{Main_HBC}(d).  This shows that the results in Fig.~\ref{Main_HBC}, which were obtained for experimentally realistic open boundary conditions and much shorter length $N_x = 50$, accurately capture the frequencies at which the delocalized bulk states are supposed to occur.  In Fig.~\ref{Fig:3}(d), we plot the range of critical frequencies versus disorder strength $W$ using the idealized (periodic boundary conditions and large $N_x$) lattice.  It displays the same levitation and annihilation behavior as in Fig.~\ref{Fig:3}(b).  The large error bar near the annihilation point ($W\approx 9J$) is due to the fact that the two humps seen in Fig.~\ref{Fig:3}(c) merge into a single large hump near the annihilation point, for the range of $N_y$ considered here.  In Fig.~\ref{Fig:3}(b), this was not observed due to the smaller lattice size and the effects of the open boundary conditions.

\begin{figure}
  \centering
  \includegraphics[width=0.475\textwidth]{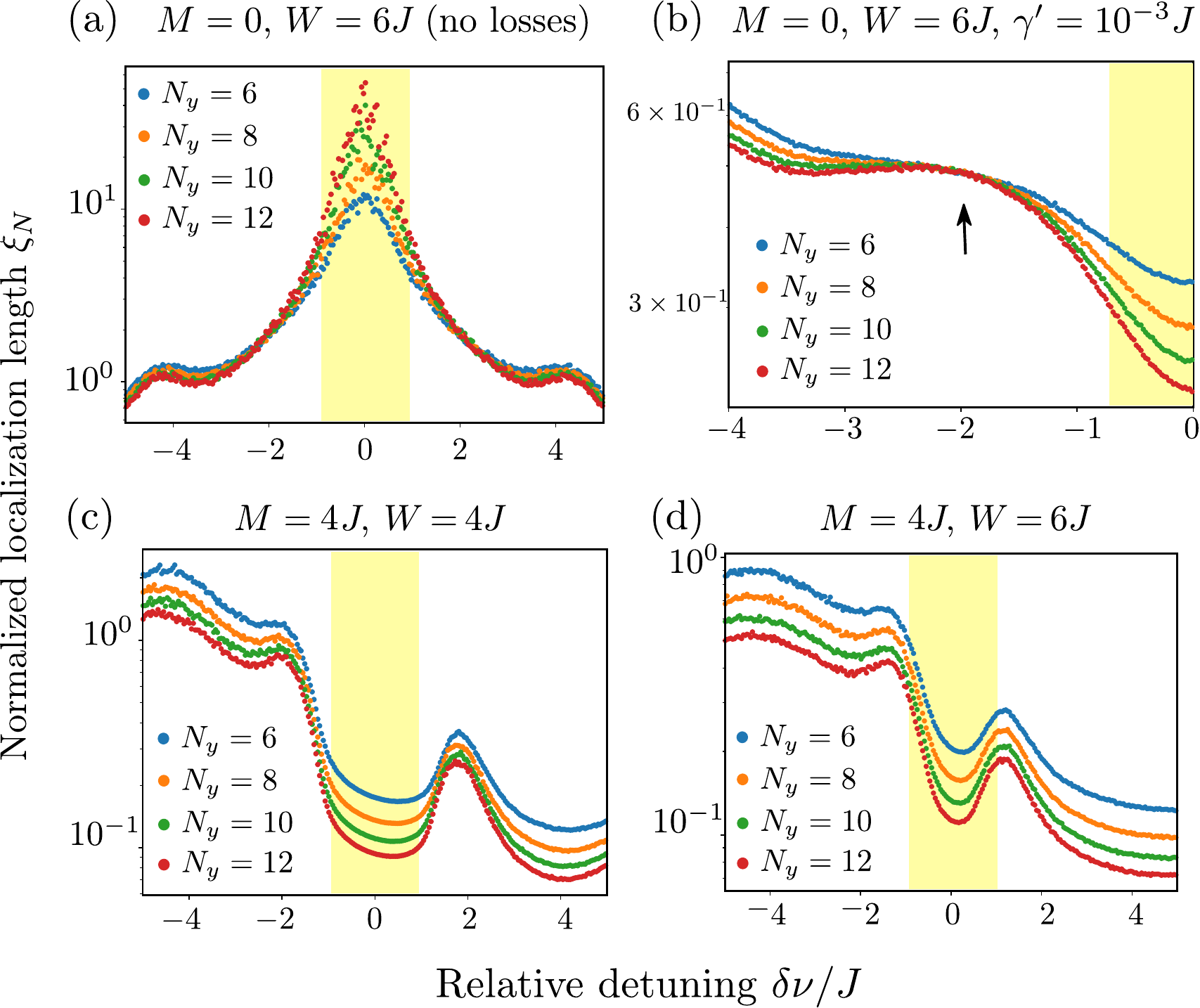}
  \caption{Plots of normalized localization length versus relative detuning $\delta \nu$ for different scenarios.  (a) Chern insulator ($M = 0$) with disorder strength $W = 6J$, with the losses on the upper and lower edge sites omitted.  Transmission now peaks in the bulk gap (yellow region) due to the topological edge states, and the signature of bulk delocalization cannot be discerned.  (b) Chern insulator ($M = 0$) with disorder strength $W = 6J$, and with small losses $-i\gamma'$ on the non-edge sites.  Here we take $\gamma' = 10^{-3}J$.  As before, the sites on the upper and lower edges have loss $-iJ$, and all other parameters are the same as in Fig.~\ref{Main_HBC}.  Bulk delocalization can be observed at $\delta \nu \approx -2 J$.  For clarity, only negative values of $\delta \nu$ are plotted.  (c)--(d) Normal insulator ($M = 4J$) with disorder strengths of (c) $4J$ and (d) $6J$ and length $N_x = 50$.}
  \label{comparisons}
\end{figure} 

To verify that the loss on the edge sites is necessary for the observation of bulk delocalization, in Fig.~\ref{comparisons}(a) we plot $\xi_N$ versus $\delta \nu$ with these losses omitted.  In this case, high transmission is observed in the frequency range corresponding to the bulk gap, due to transport by the now-unsuppressed topological edge states.  The frequency range over which $\xi_N$ is constant or increasing with $N_y$ appears to occur near the center of the bulk gap, rather than within each band as in Fig.~\ref{Main_HBC}.

In Fig.~\ref{comparisons}(b), we show that the delocalization signature can still be observed when there is weak but nonzero losses on the other resonators.  In real photonic structures, some material and radiative loss is always present.  Here, we assign each non-edge site a loss of $-i\gamma'$, where $\gamma' = 10^{-3} J$.  This level of losses is consistent what can be achieved experimentally. For example, the silicon-on-insulator implementation of the model in Ref.~\onlinecite{Mittal2019} had a loss of $\gamma^\prime \approx 0.03J$ and a silicon nitride platform \cite{Bauters2011} can achieve a loss-level smaller by a factor of around $1/100$. Therefore, a loss level of $\gamma' = 10^{-3} J$ should be achievable, and it should hence be possible to observe a frequency region in which $\xi_N$ is approximately constant with $N_y$. 

Finally, to verify the topological origin of the bulk extended state, Fig.~\ref{comparisons}(c)--(d) shows the results for $M = 4J$, for which the lattice is in its normal insulator phase.  In this case, there is no clear range of frequencies in which $\xi_N$ is constant or increasing with $N_y$.

\section{Discussion}

We have proposed an experimentally feasible way to probe the localization behavior of disordered Chern insulators using a recently-developed photonic platform \cite{Mittal2019}.  Using a realistic tight-binding model \cite{Daniel2018}, we showed that it should be possible to observe the existence of extended states in the bulk bands, a characteristic feature of disordered Chern insulators, as well as the levitation and annihilation of these extended states under increasing disorder \cite{Khmelnitskii1984, Laughlin1984, Evangelou1995, Tomi2002, Onoda_QSH2007}.  The required system sizes, disorder strengths, and loss levels are all in the experimentally accessible range.

There may be other ways to use photonic Chern insulators to probe the interplay of disorder and topological phases, such as level-spacing statistics \cite{Evers2008Rev, Castro2016}.  Apart from observing localization lengths from averaged intensity measurements, it may also be possible to probe the system using the Wigner time delay, which is insensitive to losses and has previously been used to establish the extended nature of photonic  topological edge states \cite{Mittal2014}; however, we have thus far been unable to find a clear signature of bulk state delocalization in Wigner time delay statistics.  It may also be interesting to explore the effects of adding optical gain to such a system, to determine whether bulk extended modes could be observed through their promotion into lasing modes.

Finally, although our proposal has focused on photonic resonator lattices, similar ideas can be generalized to other photonic lattices, and more broadly to other bosonic systems, such as acoustic, electrical or mechanical lattices \cite{Yang2015, Lee2018, Susstrunk2015}.

\begin{acknowledgments}
We thank Prof.~Caio Lewenkopf for helpful discussions. CYD and UC acknowledge support from the Singapore Ministry of Education Tier 3 grant MOE2016-T3-1-006 and Tier 1 grant RG187/18.  MF and SM acknowledge support from the US Air Force Office of Scientific Research (AFOSR) Multidisciplinary University Research Ini-tiative (MURI) grants FA95501610323 and FA95502010223, the US Office of Naval Research (ONR) MURI grant N00014-20-1-2325, and the National Science Foundation grant PHY1820938.
\end{acknowledgments}

\appendix

\begin{figure}[b]
  \centering
  \includegraphics[width=0.475\textwidth]{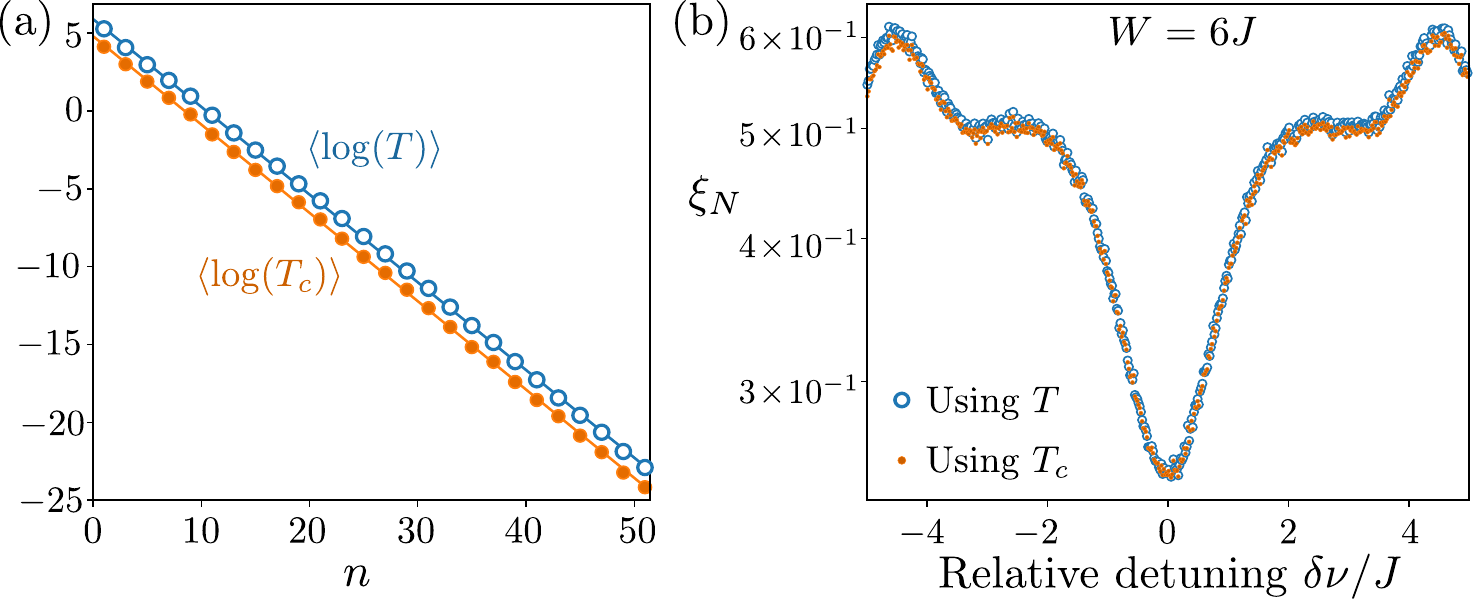}
  \caption{(a) Plots of $\langle \log (T) \rangle$ and $\langle \log (T_c) \rangle$ versus column index $n$ for samples of length $N_x = 50$, width $N_y=10$, and disorder strength $W=6J$, averaging over 500 disorder realizations.  The straight lines are linear least squares fits.  The fitted slopes are almost identical, so incoherent and coherent excitation give the same localization length estimate. (b) Normalized localization length $\xi_N$ extracted using $T$ and $T_c$, plotted against relative detuning $\delta\nu/J$ for disorder strength $W = 6J$.  All other parameters are the same as in Fig.~\ref{Main_HBC}.}
  \label{T_Tc_scaling}
\end{figure} 

\section{Coherent versus Incoherent Excitation}
\label{AppendixA}

Let $\mathcal{G}_{ba}$ denote the Green's function matrix element between site
$a$ on the input column $1$ to site $b$ on column $n$.  If the input on site $a$ is $\psi^{(\mathrm{1})}_a$, the complex wave amplitude on site $b$ is $\psi_b^{(n)} = \sum_a \mathcal{G}_{ba} \psi^{(1)}_a$, and the intensity on that site is
\begin{equation}
  I_b^{(n)} = \left|\psi_b^{(n)}\right|^2
  = \sum_{ac} G_{ba} G_{bc}^* \psi^{(\mathrm{in})}_a \psi^{(\mathrm{in})*}_c.
  \label{gsum}
\end{equation}
For uniform incoherent excitation (i.e., no fixed phase relationship between different input sites), we take an average over an ensemble of input wave amplitudes with equal magnitude $\sqrt{I_{\mathrm{in}}/N_y}$ and random phases.  Averaging over the ensemble gives the mean intensity
\begin{equation}
  \left\langle I_b^{(n)} \right\rangle = \frac{I_{\mathrm{in}}}{N_y} \sum_{a} \left|G_{ba}\right|^2,
\end{equation}
Hence, the total intensity in column $n$ is
\begin{equation}
  I^{(n)} = \sum_b \left\langle I_b^{(n)} \right\rangle =
  \frac{I_{\mathrm{in}}}{N_y} \;
  \mathrm{Tr}\left(\mathcal{G}_n^\dagger\mathcal{G}_n\right).
\end{equation}
Normalizing by the total input intensity $I_{\mathrm{in}}$ yields the transmittance formula given in Eq.~\eqref{Tstandard}.  On the other hand, if the input is spatially coherent, we do not perform an average over the random phases, and the intensity depends explicitly on the input vector, leading to Eq.~\eqref{Tcoherent}.

As shown in Fig.~\ref{T_Tc_scaling}(a), $\langle \log (T) \rangle$ and $\langle \log (T_c) \rangle$ exhibit the same scaling with distance $n$.  Hence, the same estimate for the localization length $\xi_N$ is obtained regardless of whether $T$ or $T_c$ is used in Eq.~\eqref{Trans}.  As an example, Fig.~\ref{T_Tc_scaling}(b) shows the plot of $\xi_N$ versus $\delta\nu$, similar to Fig.~\ref{Main_HBC}, for disorder strength $W = 6J$.  Estimating the localization length using either $T$ (incoherent excitation) or $T_c$ (coherent excitation) gives essentially the same results.

\bibliography{Localization_bib}

\begin{thebibliography}{60}
\expandafter\ifx\csname natexlab\endcsname\relax\def\natexlab#1{#1}\fi
\expandafter\ifx\csname bibnamefont\endcsname\relax
  \def\bibnamefont#1{#1}\fi
\expandafter\ifx\csname bibfnamefont\endcsname\relax
  \def\bibfnamefont#1{#1}\fi
\expandafter\ifx\csname citenamefont\endcsname\relax
  \def\citenamefont#1{#1}\fi
\expandafter\ifx\csname url\endcsname\relax
  \def\url#1{\texttt{#1}}\fi
\expandafter\ifx\csname urlprefix\endcsname\relax\def\urlprefix{URL }\fi
\providecommand{\bibinfo}[2]{#2}
\providecommand{\eprint}[2][]{\url{#2}}

\bibitem[{\citenamefont{von Klitzing et~al.}(1980)\citenamefont{von Klitzing,
  Dorda, and Pepper}}]{Klitzing1980}
\bibinfo{author}{\bibfnamefont{K.}~\bibnamefont{von Klitzing}},
  \bibinfo{author}{\bibfnamefont{G.}~\bibnamefont{Dorda}}, \bibnamefont{and}
  \bibinfo{author}{\bibfnamefont{M.}~\bibnamefont{Pepper}},
  \bibinfo{journal}{Phys. Rev. Lett.} \textbf{\bibinfo{volume}{45}},
  \bibinfo{pages}{494} (\bibinfo{year}{1980}).

\bibitem[{\citenamefont{Prange and Girvin}(1990)}]{prange_girvin_1990}
\bibinfo{author}{\bibfnamefont{R.~E.} \bibnamefont{Prange}} \bibnamefont{and}
  \bibinfo{author}{\bibfnamefont{S.~M.} \bibnamefont{Girvin}},
  \emph{\bibinfo{title}{The Quantum Hall Effect}}
  (\bibinfo{publisher}{Springer-Verlag}, \bibinfo{year}{1990}).

\bibitem[{\citenamefont{Wysokinski}(2000)}]{Wysokinski_2000}
\bibinfo{author}{\bibfnamefont{K.~I.} \bibnamefont{Wysokinski}},
  \bibinfo{journal}{European Journal of Physics} \textbf{\bibinfo{volume}{21}},
  \bibinfo{pages}{535} (\bibinfo{year}{2000}).

\bibitem[{\citenamefont{Anderson}(1958)}]{Anderson1958}
\bibinfo{author}{\bibfnamefont{P.~W.} \bibnamefont{Anderson}},
  \bibinfo{journal}{Phys. Rev.} \textbf{\bibinfo{volume}{109}},
  \bibinfo{pages}{1492} (\bibinfo{year}{1958}).

\bibitem[{\citenamefont{Abrahams et~al.}(1979)\citenamefont{Abrahams, Anderson,
  Licciardello, and Ramakrishnan}}]{Abrahams1979}
\bibinfo{author}{\bibfnamefont{E.}~\bibnamefont{Abrahams}},
  \bibinfo{author}{\bibfnamefont{P.~W.} \bibnamefont{Anderson}},
  \bibinfo{author}{\bibfnamefont{D.~C.} \bibnamefont{Licciardello}},
  \bibnamefont{and} \bibinfo{author}{\bibfnamefont{T.~V.}
  \bibnamefont{Ramakrishnan}}, \bibinfo{journal}{Phys. Rev. Lett.}
  \textbf{\bibinfo{volume}{42}}, \bibinfo{pages}{673} (\bibinfo{year}{1979}).

\bibitem[{\citenamefont{Evers and Mirlin}(2008)}]{Evers2008Rev}
\bibinfo{author}{\bibfnamefont{F.}~\bibnamefont{Evers}} \bibnamefont{and}
  \bibinfo{author}{\bibfnamefont{A.~D.} \bibnamefont{Mirlin}},
  \bibinfo{journal}{Rev. Mod. Phys.} \textbf{\bibinfo{volume}{80}},
  \bibinfo{pages}{1355} (\bibinfo{year}{2008}).

\bibitem[{\citenamefont{Garc\'{\i}a-Mart\'{\i}nez
  et~al.}(2013)\citenamefont{Garc\'{\i}a-Mart\'{\i}nez, Grushin, Neupert,
  Valenzuela, and Castro}}]{Garcia2013}
\bibinfo{author}{\bibfnamefont{N.~A.} \bibnamefont{Garc\'{\i}a-Mart\'{\i}nez}},
  \bibinfo{author}{\bibfnamefont{A.~G.} \bibnamefont{Grushin}},
  \bibinfo{author}{\bibfnamefont{T.}~\bibnamefont{Neupert}},
  \bibinfo{author}{\bibfnamefont{B.}~\bibnamefont{Valenzuela}},
  \bibnamefont{and} \bibinfo{author}{\bibfnamefont{E.~V.}
  \bibnamefont{Castro}}, \bibinfo{journal}{Phys. Rev. B}
  \textbf{\bibinfo{volume}{88}}, \bibinfo{pages}{245123}
  (\bibinfo{year}{2013}).

\bibitem[{\citenamefont{MacKinnon and Kramer}(1981)}]{MacKinnon1981}
\bibinfo{author}{\bibfnamefont{A.}~\bibnamefont{MacKinnon}} \bibnamefont{and}
  \bibinfo{author}{\bibfnamefont{B.}~\bibnamefont{Kramer}},
  \bibinfo{journal}{Phys. Rev. Lett.} \textbf{\bibinfo{volume}{47}},
  \bibinfo{pages}{1546} (\bibinfo{year}{1981}).

\bibitem[{\citenamefont{MacKinnon and Kramer}(1983)}]{MacKinnon1983}
\bibinfo{author}{\bibfnamefont{A.}~\bibnamefont{MacKinnon}} \bibnamefont{and}
  \bibinfo{author}{\bibfnamefont{B.}~\bibnamefont{Kramer}},
  \bibinfo{journal}{Zeitschrift f{\"{u}}r Physik B Condensed Matter}
  \textbf{\bibinfo{volume}{53}}, \bibinfo{pages}{1} (\bibinfo{year}{1983}).

\bibitem[{\citenamefont{Khmelnitskii}(1984)}]{Khmelnitskii1984}
\bibinfo{author}{\bibfnamefont{D.~E.} \bibnamefont{Khmelnitskii}},
  \bibinfo{journal}{Phys. Lett. A} \textbf{\bibinfo{volume}{106}},
  \bibinfo{pages}{182} (\bibinfo{year}{1984}).

\bibitem[{\citenamefont{Laughlin}(1984)}]{Laughlin1984}
\bibinfo{author}{\bibfnamefont{R.~B.} \bibnamefont{Laughlin}},
  \bibinfo{journal}{Phys. Rev. Lett.} \textbf{\bibinfo{volume}{52}},
  \bibinfo{pages}{2304} (\bibinfo{year}{1984}).

\bibitem[{\citenamefont{Evangelou}(1995)}]{Evangelou1995}
\bibinfo{author}{\bibfnamefont{S.~N.} \bibnamefont{Evangelou}},
  \bibinfo{journal}{Phys. Rev. Lett.} \textbf{\bibinfo{volume}{75}},
  \bibinfo{pages}{2550} (\bibinfo{year}{1995}).

\bibitem[{\citenamefont{Asada et~al.}(2002)\citenamefont{Asada, Slevin, and
  Ohtsuki}}]{Tomi2002}
\bibinfo{author}{\bibfnamefont{Y.}~\bibnamefont{Asada}},
  \bibinfo{author}{\bibfnamefont{K.}~\bibnamefont{Slevin}}, \bibnamefont{and}
  \bibinfo{author}{\bibfnamefont{T.}~\bibnamefont{Ohtsuki}},
  \bibinfo{journal}{Phys. Rev. Lett.} \textbf{\bibinfo{volume}{89}},
  \bibinfo{pages}{256601} (\bibinfo{year}{2002}).

\bibitem[{\citenamefont{Onoda et~al.}(2007)\citenamefont{Onoda, Avishai, and
  Nagaosa}}]{Onoda_QSH2007}
\bibinfo{author}{\bibfnamefont{M.}~\bibnamefont{Onoda}},
  \bibinfo{author}{\bibfnamefont{Y.}~\bibnamefont{Avishai}}, \bibnamefont{and}
  \bibinfo{author}{\bibfnamefont{N.}~\bibnamefont{Nagaosa}},
  \bibinfo{journal}{Phys. Rev. Lett.} \textbf{\bibinfo{volume}{98}},
  \bibinfo{pages}{076802} (\bibinfo{year}{2007}).

\bibitem[{\citenamefont{Mare{\u{s}} et~al.}(1999)\citenamefont{Mare{\u{s}},
  Kri{\u{s}}tofik, and Hub{\'{i}}k}}]{Mares1999}
\bibinfo{author}{\bibfnamefont{J.~J.} \bibnamefont{Mare{\u{s}}}},
  \bibinfo{author}{\bibfnamefont{J.}~\bibnamefont{Kri{\u{s}}tofik}},
  \bibnamefont{and}
  \bibinfo{author}{\bibfnamefont{P.}~\bibnamefont{Hub{\'{i}}k}},
  \bibinfo{journal}{Phys. Rev. Lett.} \textbf{\bibinfo{volume}{82}},
  \bibinfo{pages}{4699} (\bibinfo{year}{1999}).

\bibitem[{\citenamefont{Coleridge}(2005)}]{Coleridge2005}
\bibinfo{author}{\bibfnamefont{P.~T.} \bibnamefont{Coleridge}}, in
  \emph{\bibinfo{booktitle}{Encyclopedia of Condensed Matter Physics}}, edited
  by \bibinfo{editor}{\bibfnamefont{F.}~\bibnamefont{Bassani}},
  \bibinfo{editor}{\bibfnamefont{G.~L.} \bibnamefont{Liedl}}, \bibnamefont{and}
  \bibinfo{editor}{\bibfnamefont{P.}~\bibnamefont{Wyder}}
  (\bibinfo{publisher}{Elsevier}, \bibinfo{year}{2005}), p.
  \bibinfo{pages}{248}, ISBN \bibinfo{isbn}{978-0-12-369401-0}.

\bibitem[{\citenamefont{Xu et~al.}(2012)\citenamefont{Xu, Sheng, Xing, Prodan,
  and Sheng}}]{Xu2012}
\bibinfo{author}{\bibfnamefont{Z.}~\bibnamefont{Xu}},
  \bibinfo{author}{\bibfnamefont{L.}~\bibnamefont{Sheng}},
  \bibinfo{author}{\bibfnamefont{D.~Y.} \bibnamefont{Xing}},
  \bibinfo{author}{\bibfnamefont{E.}~\bibnamefont{Prodan}}, \bibnamefont{and}
  \bibinfo{author}{\bibfnamefont{D.~N.} \bibnamefont{Sheng}},
  \bibinfo{journal}{Phys. Rev. B} \textbf{\bibinfo{volume}{85}},
  \bibinfo{pages}{075115} (\bibinfo{year}{2012}).

\bibitem[{\citenamefont{Castro et~al.}(2015)\citenamefont{Castro,
  L\'opez-Sancho, and Vozmediano}}]{Castro2015}
\bibinfo{author}{\bibfnamefont{E.~V.} \bibnamefont{Castro}},
  \bibinfo{author}{\bibfnamefont{M.~P.} \bibnamefont{L\'opez-Sancho}},
  \bibnamefont{and} \bibinfo{author}{\bibfnamefont{M.~A.~H.}
  \bibnamefont{Vozmediano}}, \bibinfo{journal}{Phys. Rev. B}
  \textbf{\bibinfo{volume}{92}}, \bibinfo{pages}{085410}
  (\bibinfo{year}{2015}).

\bibitem[{\citenamefont{Qiao et~al.}(2016)\citenamefont{Qiao, Han, Zhang, Wang,
  Deng, Jiang, Yang, Wang, and Niu}}]{Qiao2016}
\bibinfo{author}{\bibfnamefont{Z.}~\bibnamefont{Qiao}},
  \bibinfo{author}{\bibfnamefont{Y.}~\bibnamefont{Han}},
  \bibinfo{author}{\bibfnamefont{L.}~\bibnamefont{Zhang}},
  \bibinfo{author}{\bibfnamefont{K.}~\bibnamefont{Wang}},
  \bibinfo{author}{\bibfnamefont{X.}~\bibnamefont{Deng}},
  \bibinfo{author}{\bibfnamefont{H.}~\bibnamefont{Jiang}},
  \bibinfo{author}{\bibfnamefont{S.~A.} \bibnamefont{Yang}},
  \bibinfo{author}{\bibfnamefont{J.}~\bibnamefont{Wang}}, \bibnamefont{and}
  \bibinfo{author}{\bibfnamefont{Q.}~\bibnamefont{Niu}},
  \bibinfo{journal}{Phys. Rev. Lett.} \textbf{\bibinfo{volume}{117}},
  \bibinfo{pages}{056802} (\bibinfo{year}{2016}).

\bibitem[{\citenamefont{Ozawa et~al.}(2019)\citenamefont{Ozawa, Price, Amo,
  Goldman, Hafezi, Lu, Rechtsman, Schuster, Simon, Zilberberg
  et~al.}}]{Ozawa2019}
\bibinfo{author}{\bibfnamefont{T.}~\bibnamefont{Ozawa}},
  \bibinfo{author}{\bibfnamefont{H.~M.} \bibnamefont{Price}},
  \bibinfo{author}{\bibfnamefont{A.}~\bibnamefont{Amo}},
  \bibinfo{author}{\bibfnamefont{N.}~\bibnamefont{Goldman}},
  \bibinfo{author}{\bibfnamefont{M.}~\bibnamefont{Hafezi}},
  \bibinfo{author}{\bibfnamefont{L.}~\bibnamefont{Lu}},
  \bibinfo{author}{\bibfnamefont{M.~C.} \bibnamefont{Rechtsman}},
  \bibinfo{author}{\bibfnamefont{D.}~\bibnamefont{Schuster}},
  \bibinfo{author}{\bibfnamefont{J.}~\bibnamefont{Simon}},
  \bibinfo{author}{\bibfnamefont{O.}~\bibnamefont{Zilberberg}},
  \bibnamefont{et~al.}, \bibinfo{journal}{Rev. Mod. Phys.}
  \textbf{\bibinfo{volume}{91}}, \bibinfo{pages}{015006}
  (\bibinfo{year}{2019}).

\bibitem[{\citenamefont{Kim et~al.}(2020)\citenamefont{Kim, Jacob, and
  Rho}}]{Kim2020}
\bibinfo{author}{\bibfnamefont{M.}~\bibnamefont{Kim}},
  \bibinfo{author}{\bibfnamefont{Z.}~\bibnamefont{Jacob}}, \bibnamefont{and}
  \bibinfo{author}{\bibfnamefont{J.}~\bibnamefont{Rho}},
  \bibinfo{journal}{Light: Science {\&} Applications}
  \textbf{\bibinfo{volume}{9}}, \bibinfo{pages}{130} (\bibinfo{year}{2020}).

\bibitem[{\citenamefont{Raghu and Haldane}(2008)}]{Raghu2008}
\bibinfo{author}{\bibfnamefont{S.}~\bibnamefont{Raghu}} \bibnamefont{and}
  \bibinfo{author}{\bibfnamefont{F.~D.~M.} \bibnamefont{Haldane}},
  \bibinfo{journal}{Phys. Rev. A} \textbf{\bibinfo{volume}{78}},
  \bibinfo{pages}{033834} (\bibinfo{year}{2008}).

\bibitem[{\citenamefont{Wang et~al.}(2009)\citenamefont{Wang, Chong,
  Joannopoulos, and Solja{\v{c}}i{\'{c}}}}]{Wang2009}
\bibinfo{author}{\bibfnamefont{Z.}~\bibnamefont{Wang}},
  \bibinfo{author}{\bibfnamefont{Y.}~\bibnamefont{Chong}},
  \bibinfo{author}{\bibfnamefont{J.~D.} \bibnamefont{Joannopoulos}},
  \bibnamefont{and}
  \bibinfo{author}{\bibfnamefont{M.}~\bibnamefont{Solja{\v{c}}i{\'{c}}}},
  \bibinfo{journal}{Nature} \textbf{\bibinfo{volume}{461}},
  \bibinfo{pages}{772} (\bibinfo{year}{2009}).

\bibitem[{\citenamefont{Rechtsman et~al.}(2013)\citenamefont{Rechtsman, Zeuner,
  Plotnik, Lumer, Podolsky, Dreisow, Nolte, Segev, and
  Szameit}}]{rechtsman2013}
\bibinfo{author}{\bibfnamefont{M.~C.} \bibnamefont{Rechtsman}},
  \bibinfo{author}{\bibfnamefont{J.~M.} \bibnamefont{Zeuner}},
  \bibinfo{author}{\bibfnamefont{Y.}~\bibnamefont{Plotnik}},
  \bibinfo{author}{\bibfnamefont{Y.}~\bibnamefont{Lumer}},
  \bibinfo{author}{\bibfnamefont{D.}~\bibnamefont{Podolsky}},
  \bibinfo{author}{\bibfnamefont{F.}~\bibnamefont{Dreisow}},
  \bibinfo{author}{\bibfnamefont{S.}~\bibnamefont{Nolte}},
  \bibinfo{author}{\bibfnamefont{M.}~\bibnamefont{Segev}}, \bibnamefont{and}
  \bibinfo{author}{\bibfnamefont{A.}~\bibnamefont{Szameit}},
  \bibinfo{journal}{Nature} \textbf{\bibinfo{volume}{496}},
  \bibinfo{pages}{196} (\bibinfo{year}{2013}).

\bibitem[{\citenamefont{Hafezi et~al.}(2011)\citenamefont{Hafezi, Demler,
  Lukin, and Taylor}}]{Hafezi2011}
\bibinfo{author}{\bibfnamefont{M.}~\bibnamefont{Hafezi}},
  \bibinfo{author}{\bibfnamefont{E.~A.} \bibnamefont{Demler}},
  \bibinfo{author}{\bibfnamefont{M.~D.} \bibnamefont{Lukin}}, \bibnamefont{and}
  \bibinfo{author}{\bibfnamefont{J.~M.} \bibnamefont{Taylor}},
  \bibinfo{journal}{Nat. Phys.} \textbf{\bibinfo{volume}{7}},
  \bibinfo{pages}{907} (\bibinfo{year}{2011}).

\bibitem[{\citenamefont{Hafezi et~al.}(2013)\citenamefont{Hafezi, Mittal, Fan,
  Migdall, and Taylor}}]{Hafezi2013}
\bibinfo{author}{\bibfnamefont{M.}~\bibnamefont{Hafezi}},
  \bibinfo{author}{\bibfnamefont{S.}~\bibnamefont{Mittal}},
  \bibinfo{author}{\bibfnamefont{J.}~\bibnamefont{Fan}},
  \bibinfo{author}{\bibfnamefont{A.}~\bibnamefont{Migdall}}, \bibnamefont{and}
  \bibinfo{author}{\bibfnamefont{J.~M.} \bibnamefont{Taylor}},
  \bibinfo{journal}{Nat. Photon.} \textbf{\bibinfo{volume}{7}},
  \bibinfo{pages}{1001} (\bibinfo{year}{2013}).

\bibitem[{\citenamefont{Leykam et~al.}(2018)\citenamefont{Leykam, Mittal,
  Hafezi, and Chong}}]{Daniel2018}
\bibinfo{author}{\bibfnamefont{D.}~\bibnamefont{Leykam}},
  \bibinfo{author}{\bibfnamefont{S.}~\bibnamefont{Mittal}},
  \bibinfo{author}{\bibfnamefont{M.}~\bibnamefont{Hafezi}}, \bibnamefont{and}
  \bibinfo{author}{\bibfnamefont{Y.~D.} \bibnamefont{Chong}},
  \bibinfo{journal}{Phys. Rev. Lett.} \textbf{\bibinfo{volume}{121}},
  \bibinfo{pages}{023901} (\bibinfo{year}{2018}).

\bibitem[{\citenamefont{Mittal et~al.}(2019)\citenamefont{Mittal, Orre, Leykam,
  Chong, and Hafezi}}]{Mittal2019}
\bibinfo{author}{\bibfnamefont{S.}~\bibnamefont{Mittal}},
  \bibinfo{author}{\bibfnamefont{V.~V.} \bibnamefont{Orre}},
  \bibinfo{author}{\bibfnamefont{D.}~\bibnamefont{Leykam}},
  \bibinfo{author}{\bibfnamefont{Y.~D.} \bibnamefont{Chong}}, \bibnamefont{and}
  \bibinfo{author}{\bibfnamefont{M.}~\bibnamefont{Hafezi}},
  \bibinfo{journal}{Phys. Rev. Lett.} \textbf{\bibinfo{volume}{123}},
  \bibinfo{pages}{043201} (\bibinfo{year}{2019}).

\bibitem[{\citenamefont{Kraus et~al.}(2012)\citenamefont{Kraus, Lahini, Ringel,
  Verbin, and Zilberberg}}]{Zilberberg2012}
\bibinfo{author}{\bibfnamefont{Y.~E.} \bibnamefont{Kraus}},
  \bibinfo{author}{\bibfnamefont{Y.}~\bibnamefont{Lahini}},
  \bibinfo{author}{\bibfnamefont{Z.}~\bibnamefont{Ringel}},
  \bibinfo{author}{\bibfnamefont{M.}~\bibnamefont{Verbin}}, \bibnamefont{and}
  \bibinfo{author}{\bibfnamefont{O.}~\bibnamefont{Zilberberg}},
  \bibinfo{journal}{Phys. Rev. Lett.} \textbf{\bibinfo{volume}{109}},
  \bibinfo{pages}{106402} (\bibinfo{year}{2012}).

\bibitem[{\citenamefont{Hu et~al.}(2015)\citenamefont{Hu, Pillay, Wu, Pasek,
  Shum, and Chong}}]{Hu2015}
\bibinfo{author}{\bibfnamefont{W.}~\bibnamefont{Hu}},
  \bibinfo{author}{\bibfnamefont{J.~C.} \bibnamefont{Pillay}},
  \bibinfo{author}{\bibfnamefont{K.}~\bibnamefont{Wu}},
  \bibinfo{author}{\bibfnamefont{M.}~\bibnamefont{Pasek}},
  \bibinfo{author}{\bibfnamefont{P.~P.} \bibnamefont{Shum}}, \bibnamefont{and}
  \bibinfo{author}{\bibfnamefont{Y.~D.} \bibnamefont{Chong}},
  \bibinfo{journal}{Phys. Rev. X} \textbf{\bibinfo{volume}{5}},
  \bibinfo{pages}{011012} (\bibinfo{year}{2015}).

\bibitem[{\citenamefont{Hu et~al.}(2017)\citenamefont{Hu, Wang, Shum, and
  Chong}}]{Hu2017}
\bibinfo{author}{\bibfnamefont{W.}~\bibnamefont{Hu}},
  \bibinfo{author}{\bibfnamefont{H.}~\bibnamefont{Wang}},
  \bibinfo{author}{\bibfnamefont{P.~P.} \bibnamefont{Shum}}, \bibnamefont{and}
  \bibinfo{author}{\bibfnamefont{Y.~D.} \bibnamefont{Chong}},
  \bibinfo{journal}{Phys. Rev. B} \textbf{\bibinfo{volume}{95}},
  \bibinfo{pages}{184306} (\bibinfo{year}{2017}).

\bibitem[{\citenamefont{Mittal et~al.}(2016)\citenamefont{Mittal, Ganeshan,
  Fan, Vaezi, and Hafezi}}]{Mittal2016}
\bibinfo{author}{\bibfnamefont{S.}~\bibnamefont{Mittal}},
  \bibinfo{author}{\bibfnamefont{S.}~\bibnamefont{Ganeshan}},
  \bibinfo{author}{\bibfnamefont{J.}~\bibnamefont{Fan}},
  \bibinfo{author}{\bibfnamefont{A.}~\bibnamefont{Vaezi}}, \bibnamefont{and}
  \bibinfo{author}{\bibfnamefont{M.}~\bibnamefont{Hafezi}},
  \bibinfo{journal}{Nat. Photon.} \textbf{\bibinfo{volume}{10}},
  \bibinfo{pages}{180} (\bibinfo{year}{2016}).

\bibitem[{\citenamefont{Zilberberg et~al.}(2018)\citenamefont{Zilberberg,
  Huang, Guglielmon, Wang, Chen, Kraus, and Rechtsman}}]{Zilberberg2018}
\bibinfo{author}{\bibfnamefont{O.}~\bibnamefont{Zilberberg}},
  \bibinfo{author}{\bibfnamefont{S.}~\bibnamefont{Huang}},
  \bibinfo{author}{\bibfnamefont{J.}~\bibnamefont{Guglielmon}},
  \bibinfo{author}{\bibfnamefont{M.}~\bibnamefont{Wang}},
  \bibinfo{author}{\bibfnamefont{K.~P.} \bibnamefont{Chen}},
  \bibinfo{author}{\bibfnamefont{Y.~E.} \bibnamefont{Kraus}}, \bibnamefont{and}
  \bibinfo{author}{\bibfnamefont{M.~C.} \bibnamefont{Rechtsman}},
  \bibinfo{journal}{Nature} \textbf{\bibinfo{volume}{553}}, \bibinfo{pages}{59}
  (\bibinfo{year}{2018}).

\bibitem[{\citenamefont{Wu and Hu}(2015)}]{Wu-Hu}
\bibinfo{author}{\bibfnamefont{L.-H.} \bibnamefont{Wu}} \bibnamefont{and}
  \bibinfo{author}{\bibfnamefont{X.}~\bibnamefont{Hu}}, \bibinfo{journal}{Phys.
  Rev. Lett.} \textbf{\bibinfo{volume}{114}}, \bibinfo{pages}{223901}
  (\bibinfo{year}{2015}).

\bibitem[{\citenamefont{Ma and Shvets}(2016)}]{Ma2016}
\bibinfo{author}{\bibfnamefont{T.}~\bibnamefont{Ma}} \bibnamefont{and}
  \bibinfo{author}{\bibfnamefont{G.}~\bibnamefont{Shvets}},
  \bibinfo{journal}{New Journal of Physics} \textbf{\bibinfo{volume}{18}},
  \bibinfo{pages}{025012} (\bibinfo{year}{2016}).

\bibitem[{\citenamefont{Lu et~al.}(2013)\citenamefont{Lu, Fu, Joannopoulos, and
  Solja{\v{c}}i{\'{c}}}}]{Lu2013}
\bibinfo{author}{\bibfnamefont{L.}~\bibnamefont{Lu}},
  \bibinfo{author}{\bibfnamefont{L.}~\bibnamefont{Fu}},
  \bibinfo{author}{\bibfnamefont{J.~D.} \bibnamefont{Joannopoulos}},
  \bibnamefont{and}
  \bibinfo{author}{\bibfnamefont{M.}~\bibnamefont{Solja{\v{c}}i{\'{c}}}},
  \bibinfo{journal}{Nat. Photon.} \textbf{\bibinfo{volume}{7}},
  \bibinfo{pages}{294} (\bibinfo{year}{2013}).

\bibitem[{\citenamefont{Peano et~al.}(2016)\citenamefont{Peano, Houde,
  Marquardt, and Clerk}}]{Peano2016}
\bibinfo{author}{\bibfnamefont{V.}~\bibnamefont{Peano}},
  \bibinfo{author}{\bibfnamefont{M.}~\bibnamefont{Houde}},
  \bibinfo{author}{\bibfnamefont{F.}~\bibnamefont{Marquardt}},
  \bibnamefont{and} \bibinfo{author}{\bibfnamefont{A.~A.} \bibnamefont{Clerk}},
  \bibinfo{journal}{Phys. Rev. X} \textbf{\bibinfo{volume}{6}},
  \bibinfo{pages}{041026} (\bibinfo{year}{2016}).

\bibitem[{\citenamefont{Zhou et~al.}(2017)\citenamefont{Zhou, Wang, Leykam, and
  Chong}}]{Zhou2017}
\bibinfo{author}{\bibfnamefont{X.}~\bibnamefont{Zhou}},
  \bibinfo{author}{\bibfnamefont{Y.}~\bibnamefont{Wang}},
  \bibinfo{author}{\bibfnamefont{D.}~\bibnamefont{Leykam}}, \bibnamefont{and}
  \bibinfo{author}{\bibfnamefont{Y.~D.} \bibnamefont{Chong}},
  \bibinfo{journal}{New J. Phys.} \textbf{\bibinfo{volume}{19}},
  \bibinfo{pages}{095002} (\bibinfo{year}{2017}).

\bibitem[{\citenamefont{Harari et~al.}(2018)\citenamefont{Harari, Bandres,
  Lumer, Rechtsman, Chong, Khajavikhan, Christodoulides, and
  Segev}}]{Harari2018}
\bibinfo{author}{\bibfnamefont{G.}~\bibnamefont{Harari}},
  \bibinfo{author}{\bibfnamefont{M.~A.} \bibnamefont{Bandres}},
  \bibinfo{author}{\bibfnamefont{Y.}~\bibnamefont{Lumer}},
  \bibinfo{author}{\bibfnamefont{M.~C.} \bibnamefont{Rechtsman}},
  \bibinfo{author}{\bibfnamefont{Y.~D.} \bibnamefont{Chong}},
  \bibinfo{author}{\bibfnamefont{M.}~\bibnamefont{Khajavikhan}},
  \bibinfo{author}{\bibfnamefont{D.~N.} \bibnamefont{Christodoulides}},
  \bibnamefont{and} \bibinfo{author}{\bibfnamefont{M.}~\bibnamefont{Segev}},
  \bibinfo{journal}{Science} \textbf{\bibinfo{volume}{359}},
  \bibinfo{pages}{eaar4003} (\bibinfo{year}{2018}).

\bibitem[{\citenamefont{Bandres et~al.}(2018)\citenamefont{Bandres, Wittek,
  Harari, Parto, Ren, Segev, Christodoulides, and Khajavikhan}}]{Bandres2018}
\bibinfo{author}{\bibfnamefont{M.~A.} \bibnamefont{Bandres}},
  \bibinfo{author}{\bibfnamefont{S.}~\bibnamefont{Wittek}},
  \bibinfo{author}{\bibfnamefont{G.}~\bibnamefont{Harari}},
  \bibinfo{author}{\bibfnamefont{M.}~\bibnamefont{Parto}},
  \bibinfo{author}{\bibfnamefont{J.}~\bibnamefont{Ren}},
  \bibinfo{author}{\bibfnamefont{M.}~\bibnamefont{Segev}},
  \bibinfo{author}{\bibfnamefont{D.~N.} \bibnamefont{Christodoulides}},
  \bibnamefont{and}
  \bibinfo{author}{\bibfnamefont{M.}~\bibnamefont{Khajavikhan}},
  \bibinfo{journal}{Science} \textbf{\bibinfo{volume}{359}},
  \bibinfo{pages}{eaar4005} (\bibinfo{year}{2018}).

\bibitem[{\citenamefont{Zeng et~al.}(2020)\citenamefont{Zeng, Chattopadhyay,
  Zhu, Qiang, Li, Jin, Li, Davies, Linfield, Zhang et~al.}}]{Zeng2020}
\bibinfo{author}{\bibfnamefont{Y.}~\bibnamefont{Zeng}},
  \bibinfo{author}{\bibfnamefont{U.}~\bibnamefont{Chattopadhyay}},
  \bibinfo{author}{\bibfnamefont{B.}~\bibnamefont{Zhu}},
  \bibinfo{author}{\bibfnamefont{B.}~\bibnamefont{Qiang}},
  \bibinfo{author}{\bibfnamefont{J.}~\bibnamefont{Li}},
  \bibinfo{author}{\bibfnamefont{Y.}~\bibnamefont{Jin}},
  \bibinfo{author}{\bibfnamefont{L.}~\bibnamefont{Li}},
  \bibinfo{author}{\bibfnamefont{A.~G.} \bibnamefont{Davies}},
  \bibinfo{author}{\bibfnamefont{E.~H.} \bibnamefont{Linfield}},
  \bibinfo{author}{\bibfnamefont{B.}~\bibnamefont{Zhang}},
  \bibnamefont{et~al.}, \bibinfo{journal}{Nature}
  \textbf{\bibinfo{volume}{578}}, \bibinfo{pages}{246} (\bibinfo{year}{2020}).

\bibitem[{\citenamefont{Villeneuve et~al.}(1996)\citenamefont{Villeneuve, Fan,
  and Joannopoulos}}]{Villeneuve1996}
\bibinfo{author}{\bibfnamefont{P.~R.} \bibnamefont{Villeneuve}},
  \bibinfo{author}{\bibfnamefont{S.}~\bibnamefont{Fan}}, \bibnamefont{and}
  \bibinfo{author}{\bibfnamefont{J.~D.} \bibnamefont{Joannopoulos}},
  \bibinfo{journal}{Phys. Rev. B} \textbf{\bibinfo{volume}{54}},
  \bibinfo{pages}{7837} (\bibinfo{year}{1996}).

\bibitem[{\citenamefont{El-Ganainy et~al.}(2019)\citenamefont{El-Ganainy,
  Khajavikhan, Christodoulides, and Ozdemir}}]{ElGanainy2019}
\bibinfo{author}{\bibfnamefont{R.}~\bibnamefont{El-Ganainy}},
  \bibinfo{author}{\bibfnamefont{M.}~\bibnamefont{Khajavikhan}},
  \bibinfo{author}{\bibfnamefont{D.~N.} \bibnamefont{Christodoulides}},
  \bibnamefont{and} \bibinfo{author}{\bibfnamefont{S.~K.}
  \bibnamefont{Ozdemir}}, \bibinfo{journal}{Comm. Phys.}
  \textbf{\bibinfo{volume}{2}}, \bibinfo{pages}{37} (\bibinfo{year}{2019}).

\bibitem[{\citenamefont{Segev et~al.}(2013)\citenamefont{Segev, Silberberg, and
  Christodoulides}}]{segev2013}
\bibinfo{author}{\bibfnamefont{M.}~\bibnamefont{Segev}},
  \bibinfo{author}{\bibfnamefont{Y.}~\bibnamefont{Silberberg}},
  \bibnamefont{and} \bibinfo{author}{\bibfnamefont{D.~N.}
  \bibnamefont{Christodoulides}}, \bibinfo{journal}{Nat. Photon.}
  \textbf{\bibinfo{volume}{7}}, \bibinfo{pages}{197} (\bibinfo{year}{2013}).

\bibitem[{\citenamefont{Wang et~al.}(2015)\citenamefont{Wang, Su, Avishai,
  Meir, and Wang}}]{Wang2015}
\bibinfo{author}{\bibfnamefont{C.}~\bibnamefont{Wang}},
  \bibinfo{author}{\bibfnamefont{Y.}~\bibnamefont{Su}},
  \bibinfo{author}{\bibfnamefont{Y.}~\bibnamefont{Avishai}},
  \bibinfo{author}{\bibfnamefont{Y.}~\bibnamefont{Meir}}, \bibnamefont{and}
  \bibinfo{author}{\bibfnamefont{X.~R.} \bibnamefont{Wang}},
  \bibinfo{journal}{Phys. Rev. Lett.} \textbf{\bibinfo{volume}{114}},
  \bibinfo{pages}{096803} (\bibinfo{year}{2015}).

\bibitem[{\citenamefont{Bauters et~al.}(2011)\citenamefont{Bauters, Heck, John,
  Dai, Tien, Barton, Leinse, Heideman, Blumenthal, and Bowers}}]{Bauters2011}
\bibinfo{author}{\bibfnamefont{J.~F.} \bibnamefont{Bauters}},
  \bibinfo{author}{\bibfnamefont{M.~J.~R.} \bibnamefont{Heck}},
  \bibinfo{author}{\bibfnamefont{D.}~\bibnamefont{John}},
  \bibinfo{author}{\bibfnamefont{D.}~\bibnamefont{Dai}},
  \bibinfo{author}{\bibfnamefont{M.-C.} \bibnamefont{Tien}},
  \bibinfo{author}{\bibfnamefont{J.~S.} \bibnamefont{Barton}},
  \bibinfo{author}{\bibfnamefont{A.}~\bibnamefont{Leinse}},
  \bibinfo{author}{\bibfnamefont{R.~G.} \bibnamefont{Heideman}},
  \bibinfo{author}{\bibfnamefont{D.~J.} \bibnamefont{Blumenthal}},
  \bibnamefont{and} \bibinfo{author}{\bibfnamefont{J.~E.}
  \bibnamefont{Bowers}}, \bibinfo{journal}{Opt. Express}
  \textbf{\bibinfo{volume}{19}}, \bibinfo{pages}{3163} (\bibinfo{year}{2011}).

\bibitem[{\citenamefont{Thouless and Kirkpatrick}(1981)}]{Thouless1981}
\bibinfo{author}{\bibfnamefont{D.~J.} \bibnamefont{Thouless}} \bibnamefont{and}
  \bibinfo{author}{\bibfnamefont{S.}~\bibnamefont{Kirkpatrick}},
  \bibinfo{journal}{J. Phys. C: Solid State Phys.}
  \textbf{\bibinfo{volume}{14}}, \bibinfo{pages}{235} (\bibinfo{year}{1981}).

\bibitem[{\citenamefont{Lewenkopf and Mucciolo}(2013)}]{Lewenkopf2013}
\bibinfo{author}{\bibfnamefont{C.~H.} \bibnamefont{Lewenkopf}}
  \bibnamefont{and} \bibinfo{author}{\bibfnamefont{E.~R.}
  \bibnamefont{Mucciolo}}, \bibinfo{journal}{Journal of Computational
  Electronics} \textbf{\bibinfo{volume}{12}}, \bibinfo{pages}{203}
  (\bibinfo{year}{2013}), ISSN \bibinfo{issn}{15698025}, \eprint{1304.3934}.

\bibitem[{\citenamefont{Zhang et~al.}(2017)\citenamefont{Zhang, Yang, and
  Chang}}]{Zhang2017}
\bibinfo{author}{\bibfnamefont{S.-H.} \bibnamefont{Zhang}},
  \bibinfo{author}{\bibfnamefont{W.}~\bibnamefont{Yang}}, \bibnamefont{and}
  \bibinfo{author}{\bibfnamefont{K.}~\bibnamefont{Chang}},
  \bibinfo{journal}{Phys. Rev. B} \textbf{\bibinfo{volume}{95}},
  \bibinfo{pages}{075421} (\bibinfo{year}{2017}).

\bibitem[{\citenamefont{Slevin et~al.}(2001)\citenamefont{Slevin, Marko{\v{s}},
  and Ohtsuki}}]{Slevin2001}
\bibinfo{author}{\bibfnamefont{K.}~\bibnamefont{Slevin}},
  \bibinfo{author}{\bibfnamefont{P.}~\bibnamefont{Marko{\v{s}}}},
  \bibnamefont{and} \bibinfo{author}{\bibfnamefont{T.}~\bibnamefont{Ohtsuki}},
  \bibinfo{journal}{Phys. Rev. Lett.} \textbf{\bibinfo{volume}{86}},
  \bibinfo{pages}{3594} (\bibinfo{year}{2001}).

\bibitem[{\citenamefont{Paulin and Carpentier}(2012)}]{Paulin2012}
\bibinfo{author}{\bibfnamefont{G.}~\bibnamefont{Paulin}} \bibnamefont{and}
  \bibinfo{author}{\bibfnamefont{D.}~\bibnamefont{Carpentier}},
  \bibinfo{journal}{New J. Physics} \textbf{\bibinfo{volume}{14}},
  \bibinfo{pages}{023026} (\bibinfo{year}{2012}).

\bibitem[{\citenamefont{Su et~al.}(2016)\citenamefont{Su, Wang, Avishai, Meir,
  and Wang}}]{Su_Srep2016}
\bibinfo{author}{\bibfnamefont{Y.}~\bibnamefont{Su}},
  \bibinfo{author}{\bibfnamefont{C.}~\bibnamefont{Wang}},
  \bibinfo{author}{\bibfnamefont{Y.}~\bibnamefont{Avishai}},
  \bibinfo{author}{\bibfnamefont{Y.}~\bibnamefont{Meir}}, \bibnamefont{and}
  \bibinfo{author}{\bibfnamefont{X.~R.} \bibnamefont{Wang}},
  \bibinfo{journal}{Scientific Reports} \textbf{\bibinfo{volume}{6}},
  \bibinfo{pages}{33304} (\bibinfo{year}{2016}).

\bibitem[{\citenamefont{Castro et~al.}(2016)\citenamefont{Castro, de~Gail,
  L\'opez-Sancho, and Vozmediano}}]{Castro2016}
\bibinfo{author}{\bibfnamefont{E.~V.} \bibnamefont{Castro}},
  \bibinfo{author}{\bibfnamefont{R.}~\bibnamefont{de~Gail}},
  \bibinfo{author}{\bibfnamefont{M.~P.} \bibnamefont{L\'opez-Sancho}},
  \bibnamefont{and} \bibinfo{author}{\bibfnamefont{M.~A.~H.}
  \bibnamefont{Vozmediano}}, \bibinfo{journal}{Phys. Rev. B}
  \textbf{\bibinfo{volume}{93}}, \bibinfo{pages}{245414}
  (\bibinfo{year}{2016}).

\bibitem[{\citenamefont{Gao et~al.}(2016)\citenamefont{Gao, Gao, Shi, Yang,
  Lin, Xu, Joannopoulos, Soljačić, Chen, Lu et~al.}}]{Gao2016}
\bibinfo{author}{\bibfnamefont{F.}~\bibnamefont{Gao}},
  \bibinfo{author}{\bibfnamefont{Z.}~\bibnamefont{Gao}},
  \bibinfo{author}{\bibfnamefont{X.}~\bibnamefont{Shi}},
  \bibinfo{author}{\bibfnamefont{Z.}~\bibnamefont{Yang}},
  \bibinfo{author}{\bibfnamefont{X.}~\bibnamefont{Lin}},
  \bibinfo{author}{\bibfnamefont{H.}~\bibnamefont{Xu}},
  \bibinfo{author}{\bibfnamefont{J.~D.} \bibnamefont{Joannopoulos}},
  \bibinfo{author}{\bibfnamefont{M.}~\bibnamefont{Soljačić}},
  \bibinfo{author}{\bibfnamefont{H.}~\bibnamefont{Chen}},
  \bibinfo{author}{\bibfnamefont{L.}~\bibnamefont{Lu}}, \bibnamefont{et~al.},
  \bibinfo{journal}{Nature Communications} \textbf{\bibinfo{volume}{7}},
  \bibinfo{pages}{11619} (\bibinfo{year}{2016}).

\bibitem[{\citenamefont{Zhang et~al.}(2013)\citenamefont{Zhang, Yang, Ju,
  Sheng, Shen, Sheng, and Xing}}]{Zhang2013}
\bibinfo{author}{\bibfnamefont{Y.-F.} \bibnamefont{Zhang}},
  \bibinfo{author}{\bibfnamefont{Y.-Y.} \bibnamefont{Yang}},
  \bibinfo{author}{\bibfnamefont{Y.}~\bibnamefont{Ju}},
  \bibinfo{author}{\bibfnamefont{L.}~\bibnamefont{Sheng}},
  \bibinfo{author}{\bibfnamefont{R.}~\bibnamefont{Shen}},
  \bibinfo{author}{\bibfnamefont{D.-N.} \bibnamefont{Sheng}}, \bibnamefont{and}
  \bibinfo{author}{\bibfnamefont{D.-Y.} \bibnamefont{Xing}},
  \bibinfo{journal}{Chinese Physics B} \textbf{\bibinfo{volume}{22}},
  \bibinfo{pages}{117312} (\bibinfo{year}{2013}).

\bibitem[{\citenamefont{Fukui et~al.}(2005)\citenamefont{Fukui, Hatsugai, and
  Suzuki}}]{Fukui2005}
\bibinfo{author}{\bibfnamefont{T.}~\bibnamefont{Fukui}},
  \bibinfo{author}{\bibfnamefont{Y.}~\bibnamefont{Hatsugai}}, \bibnamefont{and}
  \bibinfo{author}{\bibfnamefont{H.}~\bibnamefont{Suzuki}},
  \bibinfo{journal}{Phys. Soc. Jpn.} \textbf{\bibinfo{volume}{74}},
  \bibinfo{pages}{1674–1677} (\bibinfo{year}{2005}).

\bibitem[{\citenamefont{Mittal et~al.}(2014)\citenamefont{Mittal, Fan, Faez,
  Migdall, Taylor, and Hafezi}}]{Mittal2014}
\bibinfo{author}{\bibfnamefont{S.}~\bibnamefont{Mittal}},
  \bibinfo{author}{\bibfnamefont{J.}~\bibnamefont{Fan}},
  \bibinfo{author}{\bibfnamefont{S.}~\bibnamefont{Faez}},
  \bibinfo{author}{\bibfnamefont{A.}~\bibnamefont{Migdall}},
  \bibinfo{author}{\bibfnamefont{J.~M.} \bibnamefont{Taylor}},
  \bibnamefont{and} \bibinfo{author}{\bibfnamefont{M.}~\bibnamefont{Hafezi}},
  \bibinfo{journal}{Phys. Rev. Lett.} \textbf{\bibinfo{volume}{113}},
  \bibinfo{pages}{087403} (\bibinfo{year}{2014}).

\bibitem[{\citenamefont{Yang et~al.}(2015)\citenamefont{Yang, Gao, Shi, Lin,
  Gao, Chong, and Zhang}}]{Yang2015}
\bibinfo{author}{\bibfnamefont{Z.}~\bibnamefont{Yang}},
  \bibinfo{author}{\bibfnamefont{F.}~\bibnamefont{Gao}},
  \bibinfo{author}{\bibfnamefont{X.}~\bibnamefont{Shi}},
  \bibinfo{author}{\bibfnamefont{X.}~\bibnamefont{Lin}},
  \bibinfo{author}{\bibfnamefont{Z.}~\bibnamefont{Gao}},
  \bibinfo{author}{\bibfnamefont{Y.}~\bibnamefont{Chong}}, \bibnamefont{and}
  \bibinfo{author}{\bibfnamefont{B.}~\bibnamefont{Zhang}},
  \bibinfo{journal}{Phys. Rev. Lett.} \textbf{\bibinfo{volume}{114}},
  \bibinfo{pages}{114301} (\bibinfo{year}{2015}).

\bibitem[{\citenamefont{Lee et~al.}(2018)\citenamefont{Lee, Imhof, Berger,
  Bayer, Brehm, Molenkamp, Kiessling, and Thomale}}]{Lee2018}
\bibinfo{author}{\bibfnamefont{C.~H.} \bibnamefont{Lee}},
  \bibinfo{author}{\bibfnamefont{S.}~\bibnamefont{Imhof}},
  \bibinfo{author}{\bibfnamefont{C.}~\bibnamefont{Berger}},
  \bibinfo{author}{\bibfnamefont{F.}~\bibnamefont{Bayer}},
  \bibinfo{author}{\bibfnamefont{J.}~\bibnamefont{Brehm}},
  \bibinfo{author}{\bibfnamefont{L.~W.} \bibnamefont{Molenkamp}},
  \bibinfo{author}{\bibfnamefont{T.}~\bibnamefont{Kiessling}},
  \bibnamefont{and} \bibinfo{author}{\bibfnamefont{R.}~\bibnamefont{Thomale}},
  \bibinfo{journal}{Communications Physics} \textbf{\bibinfo{volume}{1}},
  \bibinfo{pages}{39} (\bibinfo{year}{2018}).

\bibitem[{\citenamefont{S{\"u}sstrunk and Huber}(2015)}]{Susstrunk2015}
\bibinfo{author}{\bibfnamefont{R.}~\bibnamefont{S{\"u}sstrunk}}
  \bibnamefont{and} \bibinfo{author}{\bibfnamefont{S.~D.} \bibnamefont{Huber}},
  \bibinfo{journal}{Science} \textbf{\bibinfo{volume}{349}},
  \bibinfo{pages}{47} (\bibinfo{year}{2015}).

\end{thebibliography}

\end{document}